\documentclass[usenatbib]{mn2e}
\usepackage{graphicx,color,epsfig}

\newcommand{\be}{\begin{equation}}
\newcommand{\ee}{\end{equation}}

\newcommand{\apj}{ApJ}
\newcommand{\apjs}{ApJS}
\newcommand{\mnras}{MNRAS}
\newcommand{\aap}{A\&A}
\newcommand{\araa}{ARA\&A}
\newcommand{\apjl}{ApJL}

\newcommand{\nat}{Nature}

\def\ltsima{$\; \buildrel < \over \sim \;$}
\def\simlt{\lower.5ex\hbox{\ltsima}}
\def\gtsima{$\; \buildrel > \over \sim \;$}
\def\simgt{\lower.5ex\hbox{\gtsima}}
\def\sgra{Sgr~A$^*$}

\newcommand\ledd{{L}_{\rm Edd}}

\def\msun{{\,{\rm M}_\odot}}

\newcommand\mbh{{\,{\rm M}_{\rm bh}}}
\newcommand\fe{Fe K$\alpha$\ }

\def\del#1{{}}

\title[Two-phase model for feedback]{Two-phase model for Black Hole feeding and feedback}

\author[S. Nayakshin]{Sergei Nayakshin\\ 
Department of Physics \& Astronomy,
  University of Leicester, Leicester, LE1 7RH, UK\\
{E-mail:~} {\rm Sergei.Nayakshin@le.ac.uk}}

\begin{document}

\date{Received}

\pagerange{\pageref{firstpage}--\pageref{lastpage}} \pubyear{2008}

\maketitle

\label{firstpage}

\begin{abstract}
We study effects of AGN feedback outflows on multi-phase inter stellar medium
(ISM) of the host galaxy. We argue that SMBH growth is dominated by accretion
of dense cold clumps and filaments. AGN feedback outflows overtake the cold
medium, compress it, and trigger a powerful starburst -- a positive AGN
feedback. This predicts a statistical correlation between AGN luminosity and
star formation rate at high luminosities. Most of the outflow's kinetic energy
escapes from the bulge via low density voids. The cold phase is pushed outward
only by the ram pressure (momentum) of the outflow. The combination of the
negative and positive forms of AGN feedback leads to an $M-\sigma$ relation
similar to the result of \cite{King03}. Due to porosity of cold ISM in the
bulge, SMBH influence on the low density medium of the host galaxy is
significant even for SMBH well below the $M-\sigma$ mass. The role of SMBH
feedback in our model evolves in space and time with the ISM structure. In the
early gas rich phase, SMBH accelerates star formation in the bulge. During
later gas poor (red-and-dead) phases, SMBH feedback is mostly negative
everywhere due to scarcity of the cold ISM.
\end{abstract}


\section{Introduction}\label{sec:intro}

 We first review the current state of the analytical AGN feedback models
  in \S \ref{sec:spherical}, and their relation to the observations. We then
  discuss two important challenges to these models that arose recently from
  microphysics of shocks, observations and numerical simulations in \S
  \ref{sec:2T} and \ref{sec:multi}. The scope of this paper and a brief
  description of the solution to these challenges are discussed in \S
  \ref{sec:scope}.

\subsection{Spherically symmetric models of AGN feedback}\label{sec:spherical}

\subsubsection{Observations and energy driven feedback}

The mass $\mbh$ of supermassive black holes (SMBH) residing in the centres of
many galaxies is observed to correlate strongly with properties of the host.
For example, $\mbh \simeq 1.5\times 10^8\sigma_{200}^4\msun$
\citep{Ferrarese00,Gebhardt00}, where $\sigma_{200} = \sigma/200~{\rm
  km\,s^{-1}}$ and $\sigma$ is the one dimensional velocity dispersion of the
stars in the host, $\sigma = (GM_{\rm bulge} /2R_b)^{1/2}$, where $M_{\rm
  bulge}$ and $R_b$ are the bulge mass and the effective radius,
respectively. Furthermore, $ \mbh \simeq 1.6\times 10^{-3}M_{\rm bulge}$
\citep{Haering04}, although a more recent census of {\em classical} bulge
systems show a higher $\mbh/M_{\rm bulge}$ ratio by a factor of a few
\citep{KormendyHo13}. More recent observations show correlations of $\mbh$
with other properties of the host
\citep{Graham04,FerrareseEtal06a,CatEtal09}. Barred galaxies show under-weight
black holes \citep{Hu08,Graham08,KormendyEtal11}, possibly indicating that
SMBH growth is fuelled not by planar inflows but rather by a ``direct''
deposition of cold clouds from the bulge \citep{NPK12a}.

Pre-dating these observations, \cite{SilkRees98} envisioned that SMBH may
influence their host galaxies strongly despite being a tiny fraction of the
total mass. They showed that {\em energy-conserving} outflows from growing
SMBH could expel all the gas in the host galaxy, terminating SMBH and galaxy
growth. This model assumes that primary outflow from the SMBH does not cool
when shocked in the interaction with the ambient gas. Quantitatively, however,
the theory predicts $\mbh \propto \sigma^5$ and requires a surprisingly
inefficient coupling between the power output of SMBH and the host. Let us
write the energy passed from the outflow to the gas in the host as $\epsilon_e
\mbh c^2$. Requiring $\epsilon_e \mbh c^2 \sim f_g M_{\rm bulge} \sigma^2$,
and $f_g\sim 0.1$ is the fractional mass of the gas in the bulge, we find that
$\epsilon_e \sim 5\times 10^{-5}$ to yield $\mbh \sim 10^{-3} M_{\rm bulge}$
at $\sigma = 200$~km/s.  Such inefficiency is puzzling. For comparison, the
radiative power output of SMBH gives efficiency of the order $\epsilon_r\sim
0.1$ \citep{Shakura73}.  In fact, the recent study of \cite{KormendyHo13}
excluded pseudo-bulges and systems currently undergoing mergers from the
sample, focusing only on the classical bulges, and obtained the SMBH to bulge
mass ratio of $\sim 0.005$ rather than $\sim 0.0015$ favored by earlier
studies. This further lowers the estimate of the energy coupling between the
UFO and the bulge to $\epsilon_e \sim 2\times 10^{-5}$. In contrast, numerical
simulations reproducing the observed correlations require efficiencies of
order $\epsilon_e \sim 5\times 10^{-3}$ \citep{DiMatteo05}.

Concluding, it appears that energy-driven feedback models simply produce too
much energy in the outflow; these models must invoke, somewhat arbitrarily, a
tiny energy coupling factor to the bulge, $\epsilon_e \sim (2-5) \times
10^{-5}$. It is not clear why this factor would be constant from one galaxy to
another, and therefore why a tight correlation between $\mbh$ and $M_{\rm
  bulge}$ \citep{KormendyHo13} would exist at all in this framework.

\subsubsection{Successes of momentum and energy driven model}

\cite{King03} proposed a more detailed AGN feedback model which is able to
account naturally for most of the relevant observations to date. In this
model, SMBH outflows start from the innermost region of the accretion discs,
and escape the region at velocity comparable to the local escape velocity,
e.g., $v_{\rm out}\sim 0.1 c$. Such outflows were actually observed in quasar
PG 1211+143 \citep{PoundsEtal03a,KP03}. The outflows carry a momentum flux
$\dot M v_{\rm out} \sim \ledd/c$, comparable to the escaping radiation
momentum flux when the SMBH luminosity is at the Eddington limit,
$\ledd$. Pleasingly, the kinetic energy carried by the ultra-fast outflow
(UFO) is $(v_{\rm out}/2c) \ledd$, which is equivalent to $\epsilon \sim
5\times 10^{-3}$, naturally accounting for the empirical results of
\cite{DiMatteo05}. Further observational support for widespread existence of
and a significant power carried by the UFOs has since become available
\citep[e.g.,][]{TombesiEtal10,Tombesi2010ApJ,PoundsVaughan11a}.

The key characteristic of this model is that it operates in {\em both
  momentum-conserving and energy-conserving} regimes at different
times. \cite{King03} has shown that close to the SMBH, shocked UFO wind
suffers significant Inverse Compton (IC) losses on the AGN radiation
field. Equalling the IC cooling time to the flow time, one finds the IC
cooling radius, $R_{\rm ic} \sim 0.5$~kpc~$M_8^{1/2} \sigma_{200}$, where $M_8
= \mbh/10^8\msun$ \citep{KZP11}. The outflow is loosing most of its kinetic
energy, $(1/2) \dot M v_{\rm out}^2$, within $R\simlt R_{\rm ic}$, and is in
the momentum-conserving regime. The ambient gas in this regime is affected
mainly by the physical push from the UFO. Considering the equation of motion
for the swept up ambient gas shell, \cite{King03} derived the maximum SMBH
mass, called $M_\sigma$ mass, above which the outflow from it clears the
galaxy of the ambient gas.

We retrace the steps in this derivation in a simpler order of magnitude
approach, only considering the force balance between the momentum outflow rate
and the gravity of the ambient swept-up shell. The structure of the ambient
gas in this model follows that in a singular isothermal sphere potential
\citep[e.g., \S 4.3.3b in][]{BT08} for simplicity. For such a potential, the
one dimensional velocity dispersion is a constant independent of radius,
$\sigma = (GM_{\rm total}/2R)^{1/2}$, where $M_{\rm total}(R)$ is the total
enclosed mass including dark matter inside radius $R$ (the distance from the
centre of the galaxy). The enclosed gas mass, $M_g(R) = f_g M_{\rm total}(R) =
2 f_g \sigma^2 R/G$ is proportional to $R$. The gas density at radius $R$ for
such a potential is
\begin{equation}
\rho_g(R) = {f_g \sigma^2 \over 2\pi G R^2} \equiv f_g \rho_0(R)\;,
\label{rho0}
\end{equation}
where we defined the total potential density, $\rho_0(R)$ for convenience;
$f_g$ is the baryon fraction. Note that the initial cosmological value of
$f_g$ is $ \simeq 0.16$ \citep[cf.][]{SpergelEtal07}. Requiring the weight of
the swept-up shell at radius $R$, $W(R) = GM(R)[M_{\rm total}(R)]/ R^2 =
4f_g\sigma^4/ G$, to be equal to the momentum flux from the AGN,
\begin{equation}
W(R) = {L_{\rm Edd}\over c} = \frac{ 4 \pi G M_{\rm bh}}{\kappa}\;,
\label{pismbh}
\end{equation}
one arrives at
\begin{equation}
\mbh = M_\sigma = f_g {\kappa \sigma^4\over \pi G^2}\approx 3.6 \times 10^8
\msun\; {f_g \over 0.16}\; \sigma_{200}^4\;.
\label{msigK}
\end{equation}
This is very close to the observed relation \citep{Ferrarese00,Gebhardt00,KormendyHo13}.

In this model, once $\mbh$ exceeds $M_\sigma$, the ambient gas is pushed
further out. The outflow accelerates once the shock expands to $R\simgt R_{\rm
  ic}$, because at that point IC losses become less important and the outflow
switches over into the {\em energy-conserving} regime. This feature of the
model is essential to explaining how the UFOs, initially momentum-driven, can
then deliver a much larger push to the ambient gas in the host
\citep{King05,King10b}. The outflow velocity of the ambient cold gas
accelerated by the UFOs in the energy-driven regime in fact reach $\sim 1000$
km~s$^{-1}$, and the mass outflow rates as large as $10^3 \msun$~yr$^{-1}$,
consistent with recent observations
\citep[e.g.,][]{FergulioEtal10,SturmEtal11b,RP11a}.  This energy-driving boost
present in the \cite{King03} model also naturally accounts for the need to
boost the momentum output of the SMBH by a factor of several over the pure
momentum-driven limit as found by \cite{SilkNusser10}.

Applying similar momentum-conserving outflow logic to stellar outflows from
young massive stellar clusters, one can account for both the observed $M_{\rm
  bh}$--$\sigma$ relation for Nuclear star Clusters
\citep[see][]{McLaughlinEtal06} and the observation that NCs are
preferentially found in low mass (low $\sigma$) galaxies
\citep[][]{NayakshinEtal09b}. The model has been also used
\citep{ZubovasEtal11a} to explain the two $\sim$10 kpc scale bubbles in the
Milky Way, emitting high energy radiation, and thus presumably filled with
cosmic rays \citep{SuEtal10}. To explain the particular geometry of the
bubbles, the only adjustment to the basic \cite{King03} model required by
\cite{ZubovasEtal11a,ZN12a} has been an addition of a dense disc of molecular
gas, known as the Central Molecular Zone, found in the central $\sim$ 200 pc
of our Galaxy \citep{MorrisSerabyn96}.


\subsection{Are shocks one or two temperature?}\label{sec:2T}

All of the AGN feedback models quoted above assumed a one-temperature model
(``1T'' hereafter) for the shocked gas, so that electrons and protons share
the same temperature everywhere in the flow. This is a reasonable assumption
for dense or relatively cold plasmas, since the electron-proton energy
exchange rate due to Coulomb collisions is large in such conditions. However,
when the plasma density becomes sufficiently low and/or ion temperature
becomes larger than $T_i \simgt 10^9$~K, the electrons and ions can thermally
decouple from each other \citep{ShapiroEtal76}.

\cite{FQ12a} showed that shocked UFOs may indeed be in this second,
two-temperature (``2T'' hereafter), regime. They found that for an outflow
velocity of $0.1 c$ and $L_{Edd}=10^{46}$erg s$^{-1}$, the ion temperature is
as high as $T_i = 2.4\times 10^{10}$K but the electron temperature reaches a
maximum of $T_e \sim 3\times 10^{9}$K in the post-shock region. IC cooling for
such ``cold'' electrons is practically negligible, in the sense that $R_{\rm
  ic}$ becomes comparable to the SMBH influence radius, that is, a few
parsec. These scales are tiny by the host galaxy's standards and are thus
unimportant.

This 2T regime for the reverse shock may be of a key importance to the
UFO-based theory of AGN feedback, since then the momentum-driven regime would
disappear (and the $M_\sigma$ mass given by equation \ref{msigK}) and we would
be back to the energy-driven paradigm. Since $\epsilon_e \sim 0.005$ in this
model, we would expect SMBH mass to be only $\sim 10^{-5} M_{\rm bulge}$, a factor of a
few hundred below what is actually observed. This is therefore a significant
logical problem.

The result of \cite{FQ12a} is not a foregone conclusion. Although observations
of fast astrophysical shocks and theoretical work \citep[see \S 2.2 for
  references in][]{FQ12a} favors the $T_e \ll T_i$ case, it is still not
entirely ruled out that collective plasma physics effects transfer the energy
between the charged species faster than Coulomb collisions
\citep[e.g.,][]{Quataert98} in the UFO setting.

However, there is now some observational support for the notion that $T_e \ll
T_i$ in the reverse shocks of the UFOs. \cite{BN13a} proposed an observational
test of whether the UFO shocks are radiative or not. They calculated the
inverse Compton cooling cascade (ICCC) emission from the spherical shocks
expected in the 1T model. The resulting spectrum in $2-10$~keV X-rays looks
like a power-law out to the roll-over energy of $\sim 50-200$~keV, depending
on the outflow velocity and the shape of the optical-UV spectrum of the
quasar. \cite{BN13a} compared this theoretical ICCC spectrum with typical
spectra of AGN and argued that there is {\em currently} no evidence for the
presence of ICCC component in the later. In particular, while the spectral
shape is reasonably similar, a whole range of variability and spectral
features such as a broad \fe, and evidence for sub-pc scale X-ray obscuration
sets the ICCC spectra apart from what is actually observed.

The results of BN13a are suggestive that the reverse shock of the UFO is in
the 2T regime but not entirely conclusive for the following reasons. It is
possible that AGN with powerful enough outflows, e.g., those in which the
kinetic power of the outflow is the assumed 5\% fraction of the bolometric
luminosity of the AGN, are rare, and future observations will discover the
"missing" ICCC component. The second possible interpretation is that the
geometry of the shock is far from spherical and therefore the expected ICCC
emission is strongly diluted. In particular, in the two-phase picture of the
ambient ISM that we are advocating here, most of the solid angles as seen from
the SMBH is filled with a hot tenuous ISM. We shall argue that this component
is driven away rapidly no matter whether the reverse shock is 1T or 2T. Most
of the UFO may thus shock at radii larger than even the 1T cooling radius and
therefore be in the energy-conserving regime. This would reduce the expected
ICCC signature perhaps below detectability.

Accordingly, in this paper we {\em }assume that UFO reverse shocks are in the
energy conserving regime everywhere for simplicity, except within the cooling
radius that is typically just a few pc in the 2T case \citep{FQ12a}, a region
that we neglect in our galaxy-wide study. Our main results are unchanged if
UFO reverse shock is actually 1T and cools efficiently within a small fraction
of the bulge.

\subsection{Multi-phase ambient gas}\label{sec:multi}

Recent well resolved 3D simulations of AGN feedback by
\cite{NZ12,ZubovasEtal13a} show that the forward shock driven into the ambient
gas is unstable to a variant of Raleigh-Taylor instability provided the outer
shock cooling time is short. The instability is physically similar to the
``Vishniac instability'' previously known in the context of supernova remnant
studies \citep{Vishniac1983,MacLowEtal89}. \cite{NZ12} finds that the
compressed outer shell breaks into filaments and massive dense clumps on the
intersection of the filaments. For $\mbh \simlt M_\sigma$, e.g., when the
shocked shell stalls and re-collapse on the SMBH, the dense filaments collapse
onto the SMBH the fastest. Physically, the filaments experience a much smaller
outward acceleration due to the outflow because they have a very large column
depth compared with the mean for the shell.

\begin{figure}
\psfig{file=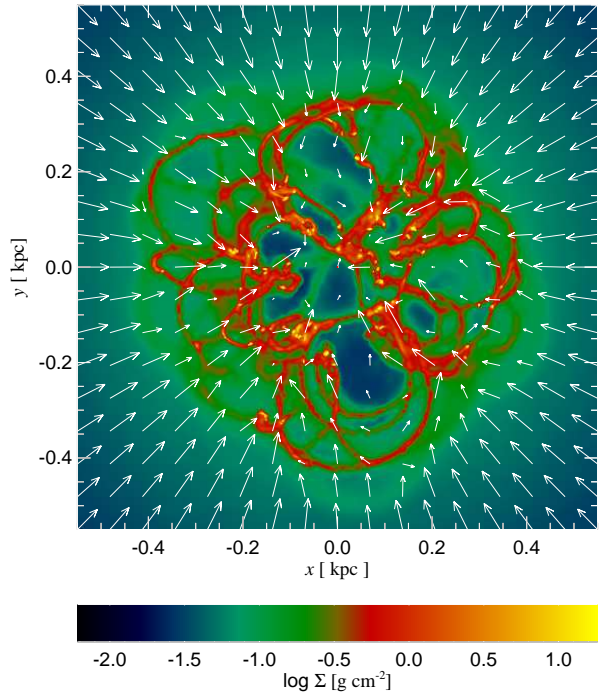,width=0.5\textwidth,angle=0}
\caption{Projected column density (colours; the scale bar is on the bottom of
  the figure) and velocity vectors of the gas for a simulation of AGN feedback
described in the text. Note the dense filaments and clumps that are infalling
onto the SMBH.}
\label{fig:sim}
\end{figure}

Figure \ref{fig:sim} presents the gas column density and velocity field for
the simulation presented in \cite{NZ12}. The SMBH mass in this simulation is
$\mbh = 10^8\msun$, which is $\sim 30$\% below a numerically found $M-\sigma$
mass for the setup \citep[see the simulations in][that show $\mbh > M_\sigma$
  cases for the same potential]{ZubovasEtal13a}. The densest parts of the
shell eventually collapse onto the black hole in this simulation, while the
lower density ``bubbles'' grow larger than in figure \ref{fig:sim}. 

This figure demonstrates very clearly that validity of the
spherically-symmetric shell approximation for AGN feedback in a situation such
as obtained here must break down. We would like to emphasise that the
simulations of \cite{NZ12,ZubovasEtal13a} were started from spherically
symmetric initial conditions. This is why the gas outside the inner $\sim 0.4$
kpc is uniform in figure \ref{fig:sim}. Gas in realistic galaxies is expected
to be multi-phase even without AGN feedback, as soon as the radiative cooling
time in the bulge is comparable to the dynamical time
\citep{McKeeOstriker77,FallRees85}. 3D numerical simulations clearly show that
thermal instabilities in the gas lead to a cooling runaway in denser regions;
these regions not only become cooler but are also further compressed by the
virialised surrounding hot gas
\citep[e.g.,][]{BaraiEtal12,HopkinsEtal12a,GaspariEtal13,MoProga13}. Such
thermal instabilities and/or turbulence induced by supernova explosions in the
bulge are likely to lead to the chaotic AGN accretion mode in which dense
filaments stream ballistically into the central parsecs of the host
\citep[e.g.,][]{HobbsEtal11}.

Besides testing the validity of the spherically symmetric models, simulations
such as that presented in Fig. 1 also indicate an important new physical
effect of AGN feedback on its host: triggered star formation in the cold dense
gas caused by the extremely high pressure of the UFO shocks
\citep[e.g.,][]{NZ12,ZubovasEtal13a,ZubovasEtal13b,Silk13}. Such a positive
AGN feedback was also invoked by \cite{SilkNusser10} to provide for extra
feedback within the bulge.

Finally, observations of AGN-driven molecular outflows constrain mean density
and the total mass of molecular gas in the host, which then implies that the
gas must reside in dense clumps \citep[e.g.,][]{AaltoEtal12,CiconeEtal11}, as
envisioned above.

\subsection{The role of galactic discs}\label{sec:disc}

Our model posits that most of SMBH growth occurs due to accretion of cold
  clumps or filaments from a quasi-spherical bulge of the galaxy. As discussed
  above, hot gas from the bulge does not contribute to SMBH growth
  significantly unless it cools and switches phases. The other likely and
  significant reservoir of cold gas is a galactic disc which must form
  inevitably if there is sufficient angular momentum in the host galaxy. 
This disc could in principle fuel SMBH growth instead of the cold clumps, but we argue that 
galaxy-scale discs do not transfer angular momentum
sufficiently rapidly to provide the dominant source of AGN feeding.

On the
theory side, this view is supported by the fact that AGN accretion discs are
unstable to self-gravitational instability beyond $\sim 0.1$~pc; it is hard to
see how such discs could fuel AGN \citep[e.g.,][]{Goodman03,NK07} for tens of
million of years, the likely duration of SMBH growth phase.

Observations also
disfavour large discs (or bar) as the dominant sources for SMBH growth.  The view developing observationally is that bright quasars reside in classical bulges or elliptical galaxies, whereas galaxies dominated by discs host AGN that are typically dimmer by one to two orders of magnitude \citep{KormendyHo13}.

Furthermore, these SMBH feeding-based objections against disc-dominated growth of SMBH can be also backed up from the AGN feedback angle. \cite{NPK12a}
showed that if galactic scale disc were able to self-regulate their star
formation rate due to stellar feedback within the discs, as argued by
\cite{Thompson05}, and thus feed the SMBH efficiently, then one would expect
SMBH in galaxies with prominent discs or bars to be a factor of a few to ten
more massive than SMBH in elliptical galaxies at same velocity
dispersion. This conclusion stems from a simple geometric argument: most of
AGN feedback misses the disc. It is thus very hard to stop SMBH growth via
a disc or a bar {\em if} those processes were efficient if feeding the
SMBH. Observations of pseudo-bulge systems show the opposite result: SMBH in
such galaxies are under-weight with respect to their cousins in classical
bulges by a factor of a few to ten \citep{Hu08,Graham08}, and in fact may not
even correlate with $\sigma$ \citep{KormendyEtal11,KormendyHo13}.

The final problem with a planar mode of SMBH feeding is pointed out by 
\cite{ZubovasEtal13b}, who showed that pressure in the hot bubbles inflated by
AGN feedback is much larger than the pressure in the self-regulated galactic
discs studied by \cite{Thompson05}. These discs are thus over-pressured by the
bubbles into much more rapid star formation than on its own. This speeds up
their transformation into stellar rather than gaseous discs and makes SMBH
feeding even harder.

For these reasons galactic discs do not feature in our paper. They are
important engines for formation of galactic stellar discs and transfer of
matter on kpc-scales, but we see no theoretical or observational basis to
connect such discs to {\em most} of SMBH growth.

\subsection{The scope and main results of the paper}\label{sec:scope}

We believe that the basic scenario for the AGN feedback proposed by
\cite{King03} is the most promising of all the models available in the
literature to date since it is based on robust physical expectations for
near-Eddington AGN accretion flows and actual observations of ultra-fast
outflows, as described in \S 1.1. The goal of our paper is to ask how this
model should be modified when one takes into account a number of the latest
results in the field: (i) the two-temperature regime for the UFO reverse
shock; (ii) the multi-phase structure of the ambient and shocked ambient gas;
(iii) AGN-triggered star formation in the host; and (iv) chaotic AGN
accretion.

We find that in a realistic two-phase environment, (a) cold ambient gas is
driven outward by the outflow's momentum only \citep[as in][model]{King03}
despite the outflow being energy conserving, and (b) the hot/low-density phase
is affected by both the energy and the momentum of the outflow, so being
driven off much easier than the cold gas; (c) star formation in the densest
clouds in the cold phase is essential in limiting SMBH growth.  The
implications of our results are reviewed in \S \ref{sec:discussion}.

\section{AGN feedback on a multi-phase ambient gas}\label{sec:model}

\subsection{The model and assumptions}\label{sec:assumptions}

\begin{figure}
\psfig{file=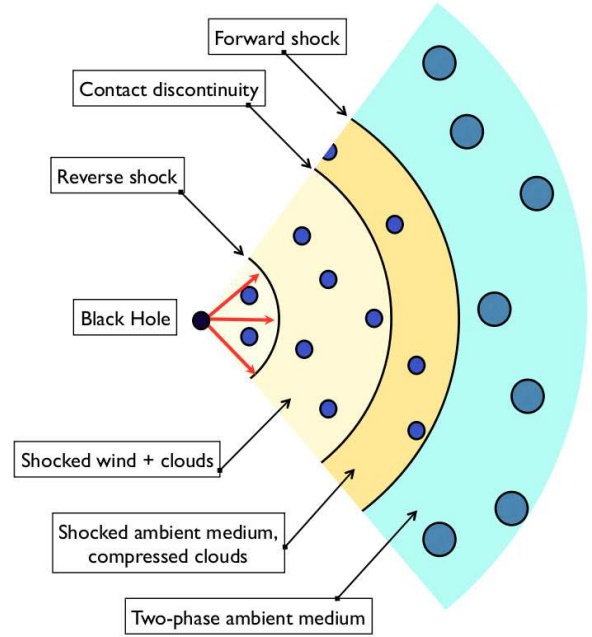,width=0.88\textwidth,angle=0}
\caption{Sketch of the geometry of the problem. The Black Hole is at the
  centre of a quasi-spherical host bulge. Ambient gas in the host consists of
  a low density hot medium enveloping a cold high density clouds. The
  ultra-fast outflow from the AGN shocks mainly against the hot low density
  gas, driving an energy-driven flow out to great distances. The cold dense
  clouds are pushed outward just by the ram pressure of the outflow but are
  also compressed by its high pressure and experience a triggered star
  formation bust. See text in \S \ref{sec:assumptions} for more detail.}
\label{fig:sketch}
\end{figure}

We follow the model of \cite{King03} in that the SMBH produces a momentum flux
during its Eddington-limited outburst of $\ledd/c$, and that the outflow
velocity is $v_{\rm out}\sim 0.1 c$. We also assume an isothermal galaxy
potential dominated by dark matter, and a quasi-spherical distribution of
ambient gas in the host. In variance to \cite{King03}, however, we assume that
the ambient medium consists of two phases -- ``hot'' and ``cold'', as sketched
in Figure \ref{fig:sketch}. The hot medium's temperature is about the virial
temperature, $k_b T_{\rm hot}/\mu \approx GM_{\rm tot}/R = 2\sigma^2$, whereas
the cold clouds are much cooler. The hot volume phase occupies most of the
volume inside the host, while the cold medium probably carries most of the
gaseous mass \citep{McKeeOstriker77,FallRees85}.

In figure \ref{fig:sketch}, arrows indicate ultra-fast outflow from the SMBH
which is located at the centre of the bulge of the host galaxy. We assume that
the hot medium interacts with the UFO as found in the spherically uniform
models \citep[e.g.,][]{King10b,ZK12a,FQ12a}, except the hot gas mass fraction
(see below) may be lower and variable with radius. Counting from the SMBH,
there is then three important surfaces where the nature of the flow changes
discontinuously: the reverse shock, the contact discontinuity and the forward
shock, all indicated in the figure. The cold medium however is able to
``penetrate'' the shocks in the sense that the shocked outflowing material
overtakes the cold clouds.

One of the points of our paper is to show that the momentum of the ultra-fast
outflow is still key to establishing the $M-\sigma$ relation even if the
outflow is always in the energy-driven regime, as suggested by the results of
\cite{FQ12a}. We therefore assume that the outflow is in the energy-conserving
mode everywhere.

\subsection{The hot phase: energy driving}\label{sec:hot}

We argue that the density of the hot ambient medium is determined by the
balance between virialisation shocks and radiative cooling. This requires that
the density of the hot gas is approximately that which gives radiative cooling
time of the order of dynamical time \citep{FallRees85}. Using the cooling
function of \cite{Sutherland93}, we find
\begin{equation}
f_{\rm h} = {\rho_{\rm h}\over \rho_0} \approx 2.5\times 10^{-3}
\sigma_{200}^{5/2} R_{\rm kpc} z^{-0.6}\;,
\label{fh1}
\end{equation}
where $z$ is the metalicity of the gas in units of Solar metalicity, $R_{\rm
  kpc}$ is distance $R$ in units of 1 kpc. This estimate would be somewhat
higher if we also considered feedback from a likely ongoing star formation in
the host, but even this does not change the main conclusion: $f_{\rm h}$ is
quite small compared with the initial cosmological gas fraction $f_0 = 0.16$,
so that most of the mass is expected to reside in the cold gas phase
\citep[also, cf. simulations by][]{HopkinsEtal12a}.

The outflow interacts with the two phases differently.  The outflow shocks
against the hot medium and drives it outward in the energy-driven regime, as
calculated by \cite{King03,King10b,ZubovasEtal11a}, except that the hot gas
density fraction $f_{\rm h}$ is lower than $f_0$. Since the energy driving
regime is much more efficient than the momentum-driven one, and since $f_{\rm
  h}$ is small, the hot medium is expelled relatively easily. To
  appreciate our point, let us consider the binding energy of the hot
  component in the bulge, writing it by order of magnitude as $E_{\rm h} \sim
  M_{\rm h} \sigma^2 = f_{\rm h} M_{\rm bulge} \sigma^2$, where $M_{\rm h} =
  f_{\rm h} M_{\rm bulge}$ is the mass of the hot ISM in the bulge of mass
  $M_{\rm bulge}$. Compare this energy to kinetic energy of the UFO emitted by
  the SMBH, $E_{\rm UFO} \sim 0.1 \mbh (v_{\rm out}^2/2)$:
\begin{equation}
{E_{\rm h}\over E_{\rm UFO} } \sim 10^{-3} \; {f_{\rm h}\over 0.01}\; \sigma_{200}^2\;.
\end{equation}
Here we used the observational fact that bulges are typically $\sim 1000$
times more massive than SMBH they host \citep{Haering04}.  Since $f_{\rm h}$
is much smaller than unity and definitely cannot exceed unity by definition we
see that $E_{\rm h}/E_{\rm UFO} \ll 1$ always, and it thus must be an easy
task for the UFO to remove the hot phase from the bulge\footnote{Of course in
  reality smaller cold clouds are destroyed by the UFO shocks, and thus there
  is a continuous replenishment of the hot phase in the bulge.}.

Furthermore, we note that SMBH is not likely to be fed directly by the hot
phase gas for at least two reasons. First of all, as is well known, hot gas
overheats inside non-radiative accretion flows and gets unbound
\citep{Blandford99}; secondly, even if hot gas could fuel some AGN activity,
its low density makes it very vulnerable to feedback outflows from AGN. Some
of the well known astrophysical examples of this behaviour are the SMBH in the
nearby giant elliptical galaxy M87 \citep[e.g.,][]{DiMatteoEtal03}, \sgra, the
SMBH in the centre of our Galaxy \citep{Baganoff03b,Cuadra06}, and ``cooling
flows'' in galaxy clusters \citep{ChurazovEtal02}.

Since we concluded that the hot ISM does not play a significant role in SMBH
feeding and is blown away in the energy-conserving mode efficiently by UFOs,
our thesis is that this phase plays no role in establishing the $M-\sigma$
relation.

\subsection{The cold phase: momentum driving}\label{sec:cold}

In our model, similarly to the interaction of supernova blast waves with cold
ambient clouds, the fast outflow from the AGN overtakes cold clouds rather
than pushes them in front of itself \citep[e.g.,][]{McKeeCowie75}. Lower
density clouds are probably crashed and dispersed, with their gas joining the
ambient hot shocked gas (thus increasing $f_{\rm h}$ from the estimate
above). Higher density clouds however survive and lag behind the forward
shock; they are eventually entrained in both the forward and the reverse shock
regions.

We note that highest density clouds are much more resilient to AGN feedback
and can be pulled inward by gravity even in the presence of an energetic
outflow (cf. fig. \ref{fig:sim}). For this reason the SMBH is fed in our model
by high density clouds rather than by the ``mean'' density ambient gas. We
therefore envisage that cold phase permeates the host galaxy everywhere (see
fig. \ref{fig:sketch}), although it fills a small fraction of the galaxy's
volume.

To gain some analytical insight into this complicated problem, let us consider
the cold phase to be a population of spherical clouds, each of mass $m$,
spherical radius $r$, and $\Sigma = m/(\pi r^2)$, the cloud's column
density. The UFO shocks against the cloud, building up a bow shock in front of
it \citep[as in figure 1 of][]{McKeeCowie75}. Importantly, considering the
radial motion of the cloud, we only need to include the momentum flux of the
fast outflow directly impacting the cloud; while the fast outflow shocks in
the bow shock region, the thermalised energy of the shock simply overflows the
cloud sideways \citep{McKeeCowie75}.

As in the \cite{King03} model, the SMBH outflow produces ram pressure
(momentum flux) at distance $R$ from the SMBH equal to
\begin{equation}
P_{\rm ram} = {\ledd \over 4\pi R^2 c} = {G\mbh \over R^2 \kappa}\;.
\label{pram1}
\end{equation}
The inward directed gravitational force acting on the cloud is balanced by the ram
pressure of the outflow when
\begin{equation}
{GM_{\rm tot}(R)\over R^2} m \sim P_{\rm ram} \pi r^2\;.
\label{feq1}
\end{equation}
Outflow's ram pressure on the cloud exceeds gravitational force from the bulge if
\begin{equation}
\mbh \simgt {\kappa \sigma^4\over \pi G^2} {\rho\over \rho_0} {r\over R}\;.
\label{ms1}
\end{equation}
However, we should also estimate the number of clouds on a line of sight,
$N_l$, as seen from the SMBH. Trivial geometrical considerations show that
this is of the order of $N_l \sim O(1) f_{\rm c} \rho_0 R/(\rho r)$, where
$f_c \equiv \rho_{\rm c}/\rho_0$ is the volume averaged cold gas mass
fraction, equivalent to $f_g$ in the model of \cite{King03}, except that the
latter encompasses all of the gas. Choosing $O(1)\sim 1$, we get the result
formally correct in the case of smooth spherically symmetric cold medium (that
is, if $f_{\rm h}=0$, we should have exactly 1 cold ``cloud'' with density
$\rho = f_{\rm c} \rho_0$ and $r=R$ per line of sight), so we write
\begin{equation}
N_l \sim  f_{\rm c} {\rho_0 R \over \rho r}\;.
\label{Nl}
\end{equation}
We then require that in general the momentum flux from the SMBH must exceed
the weight of $(1 + N_l)$ clouds. This is asymptotically correct in the
corresponding opposite limits, $N_l \ll 1$ (when we need to consider only one
cloud as other clouds do not shadow it), and $N_l\gg 1$. With this, the
critical SMBH mass in our model is
\begin{equation}
\mbh \sim {\kappa \sigma^4\over \pi G^2} \left( f_c + {\rho r \over \rho_0 R}
     \right)\;.
\label{ms2}
\end{equation}

\subsection{Star formation limiter and the $M-\sigma$ relation}\label{sec:sfr}

In principle, clouds could be arbitrarily dense, $\rho\gg \rho_0$, which would
make it all but impossible to stop them from falling in despite the ram
pressure from the UFO. Formally, in this limit the column depth of the cold
clouds, $\Sigma \sim \rho r$, could be arbitrarily high, so that for densest
clouds $\Sigma \gg \rho_0 R$ and thus the last term in equation \ref{ms2}
becomes arbitrarily large. SMBH could then grow by accretion of such clouds to
masses much larger than observed.

However, clouds that have very large densities and short cooling times are
self-gravitating and are liable to fragmentation into stars. We argue that
such clouds cannot feed the SMBH because the constituent gas is partially
turned into stars, while the remainder of the cloud is disrupted (unbound) by
star formation feedback \citep[this is the fate of most Galactic star-forming
  molecular clouds, e.g.,][]{McKeeOstriker07}. In the presence of AGN outflow
the remnants of the star-formation disrupted cloud are shocked, heated up and
mixed with the AGN outflow. 

Now consider conditions for a cloud to be self-gravitating: their self-gravity
must exceed the tidal force from the bulge, so that
\begin{equation}
{Gm\rho\over r^2}\sim {GM_{\rm tot}\over R^2} {r \over R}\;.
\end{equation}
By the order of magnitude, we can conclude from the above that
\begin{equation}
\rho_{\rm sg} \sim {\sigma^2\over 2 \pi G R^2} =  \rho_0\;.
\label{rho_sg}
\end{equation}
gives the mean density of a cloud that is just self-gravitating, where
$\rho_0$ is the density of the background potential introduced by equation
\ref{rho0}.

Therefore, the maximum density of gas clouds that fuel SMBH growth is limited
by $\rho_{\rm sg}\sim \rho_0$ in our model. The radial size of the clouds,
$r$, depends on the cloud's mass $m$, but clearly cannot be larger than some
small fraction of radius $R$. We thus introduce a parameter $\delta \equiv
\rho r/(\rho_0 R)\ll 1$, averaged over the ensemble of cold clouds in the
host. With this we write the critical ``cold'' SMBH mass as
\begin{equation}
M_{\rm cold} \sim  {\kappa \sigma^4\over \pi G^2} \left(f_{\rm c} +
\delta\right) \;= \; 2.2\times 10^8\msun
{f_{\rm c} + \delta\over 0.1} \sigma_{200}^4\;.
\label{mine1}
\end{equation}

This is the critical SMBH mass which should be compared to the observed
$M-\sigma$ relation \citep[e.g.,][]{KormendyHo13}.

\section{Discussion}\label{sec:discussion}

Observations require a successful AGN feedback theory to operate in both the
momentum-conserving regime -- to explain the $M-\sigma$ relation and the low
efficiency of AGN feedback coupling to the gas in the host, and in the
energy-conserving regime to explain the massive molecular and ionised outflows
observed \citep[e.g.,][]{FergulioEtal10,SturmEtal11b,RP11a}. In the
spherically symmetric one-phase model \citep{King03,King05,King10b} these two
regimes appear naturally. While SMBH is below the $M-\sigma$ mass, the outflow
stalls and looses most of its energy by IC radiation in the central few
hundred pc of the bulge, and is thus momentum driven in that region. Once
$\mbh > M_\sigma$, the clears this central region and it then switches over
into the energy-conserving regime.

The two-phase model proposed here is similar to that of \cite{King03} in many
regards. We use exactly same model for the UFO, which we believe is the main
culprit of AGN feedback on the host. We also find the momentum-driven and the
energy-driven regimes for the UFO interaction with the ambient medium. We
derive an $M-\sigma$ mass that is formally similar to the one obtained by
\cite{King03}. However, there are also significant differences in the
assumptions of the model: (a) most importantly, the ISM of the host galaxy is
multi-phase in our model; and (b) we assume the reverse shock to be in two
temperature (2T) regime \citep[see][]{FQ12a}, although our results are not
very sensitive to this assumption, and will not change significantly if the
electrons and ions have same temperature.

  We now summarise key conclusions and results from this paper:

\begin{itemize}

\item[(1)] Equation (\ref{mine1}) gives an $M-\sigma$ relation for cold clouds exposed to
AGN outflow carrying the momentum flux $\ledd/c$. It is similar to the result
of \cite{King03}, despite the fact that we assumed that ultra-fast outflows
are energy-conserving everywhere. Our result follows mainly from two
conclusions -- that cold dense clouds are overtaken by the UFOs, so experience
only the momentum push, and that the density of the clouds that can feed the
SMBH is limited by cloud self-gravity. Therefore, as in \cite{King03}, both
momentum and energy of the UFO are important, the former for driving the cold
clouds away whereas the latter for driving lower density gas to high outward
velocities. It is important to emphasise at this point that the ``hot shocked
phase'' may well contain cooler cloud inclusions that are denser and could
thus cool down, in principle all the way to temperatures at which molecules
form. This phase could be accelerated to high velocity and even form stars
\citep{ZubovasEtal13a}.

\item[(2)] One of the key assumptions of our model is that gas is delivered to
  the SMBH by cold dense clouds. This naturally connects to the observational
  fact that accretion of hot virialised gas on SMBH is inefficient and
  proceeds at rates much lower than the expected Bondi rate
  \citep[e.g.,][]{Baganoff03A,DiMatteoEtal03}. This inefficiency is in
  contrast to the recent ideas about the ``chaotic AGN feeding'' regime, in
  which clouds with randomly oriented angular momentum are deposited into the
  central $\sim$ parsec of the host. \cite{KingPringle06} argued that this
  helps to alleviate the problem of too large a spin of SMBHs that is not
  observed, \cite{NK07,KingPringle07} suggested that this would also solve the
  self-gravity ``catastrophe'' of AGN discs
  \citep{Goodman03,NayakshinEtal07}. \cite{HobbsEtal11} proposed that
  turbulence induced in the bulge by star formation may naturally produce such
  chaotic inflows \citep[also see][for related
    ideas]{BaraiEtal12,GaspariEtal13}.

We also suggested that density of clouds that eventually make it deep enough
to feed the SMBH is limited by their self-gravity, since denser clouds
participate in star formation instead. This makes a testable prediction: there
can be no significant SMBH growth without star formation in the host, since if
there are dense clouds then some of them are inevitably forming stars. This is
consistent with the observations: star formation and bright AGN activity are
well known to go hand in hand. For example, \cite{ChenEtal13a} report a nearly
linear correlation between star formation rates and SMBH accretion rates for a
sample of star forming galaxies with redshift $0.25 < z < 0.8$. This
correlation is only statistical however, as we expect AGN luminosity to vary
on short time scales, possibly by multiple orders of magnitude \citep[see more
  on that in][]{ZubovasEtal13b}, whereas star formation rate is not likely to
vary on time scales shorter than $\sim 1$ Myr.

\item[(3)] In spherically-symmetric one-phase models, the $M-\sigma$ relation
  arises because the SMBH drives gas outward, limiting its own growth. In our
  model this is only part of the answer; SMBH growth is limited by both the
  outward push and star formation triggered in the host by the UFO. In our
  model AGN feedback does not terminate star formation in the host as is often
  assumed, it rather accelerates it when it can (while the cold ISM phase is
  abundant). The termination step does occur but only at late times when
  there is little gas in the bulge left. In this case most of the ISM is
  in the hot phase, and it takes very little effort from the SMBH to remove it
  or stir it up to prevent cooling.

\item[(4)] The overall sign of AGN feedback is always negative, since in the
  absence of the outward forces caused by the feedback, all the gas that fell
  into the host could be eventually converted into stars. However, it is
  important in our picture that feedback of the AGN onto the galaxy's bulge
  can be both positive (triggered star formation) and negative (gas removal).

\item[(5)] It is also interesting to note that $M-\sigma$ relation is
  established inside quite a small part of the host bulge in the
  spherically-symmetric model \citep{King03}, e.g., inside $R\simlt R_{\rm
    ic}\sim 0.5$~kpc~$M_8^{1/2} \sigma_{200}$. The size of the bulge is much
  larger, e.g., $R_{\rm b} = G M_{\rm bulge}/2\sigma^2 \sim
  30$~kpc~$\sigma_{200}^2$. Those large scales are readily cleared in the
  energy driven regime in that model. This however implies that SMBH has a
  negligible effect on the host galaxy while $\mbh < M_\sigma$, since the
  momentum-driven outflows stall in the region $R\simlt R_{\rm ic}$ until the
  SMBH exceeds the $M_\sigma$ mass.  In our model, on the other hand, even
  black holes with mass $\mbh \ll M_\sigma$ are important for their hosts
  because the outflow is in the energy conserving mode and is able to
  percolate through the cold medium, therefore affecting the galaxy far and
  wide. Therefore, a key observational test of our model would be detection of
  an energy driven massive outflow from a SMBH that is well below its
  $M_\sigma$ mass. Such an outflow could not possibly exist in the homogeneous
  one-phase model for AGN feedback.

\end{itemize}

\cite{CiconeEtal11} report a massive outflow in the ULIRG hosting galaxy
  Mrk 231, carrying as much as $\dot M \sim 700 \msun$~yr$^{-1}$ of cold gas
  traced in CO and HCN molecular transitions. The outflow extends to $\sim 1$
  kpc scale and shows velocities of $\sim 1500$~km~s$^{-1}$. Such energetics
  is consistent with the energy-driven flow that entrained molecular
  clouds. Interestingly, the authors find that the extent of the outflow
  anti-correlates with the critical densities of the line transition used to
  trace the outflow. They interpret this correlation as a signature that
  density of the clouds responsible for the lines observed decrease with
  radius, indicating that high density clouds evaporate as they flow
  outwards. We note that this observation could be interpreted no less
  naturally in the context of our model. Since the pressure exerted by the UFO
  onto the clouds decreases with radius \citep[see][]{ZubovasEtal13b}, clouds'
  density must indeed fall with increasing distance from the AGN. This
  conclusion does not require cloud destruction at all.

Finally, we emphasise that the analytical model developed here is an attempt
to obtain a simple yet useful analytical insight into a very complicated
problem. One must keep in mind that in reality there is a constant mass
exchange between the phases, e.g., by cloud formation and destruction via
thermal conduction evaporation, shocks and instabilities
\citep[e.g.,][]{McKeeOstriker07,FEtal12a}. To give an example, a cold cloud
moving on a radial trajectory out of the host could have started as a much
lower density but hotter gas that was shocked and compressed by the UFO and
then cooled down, all the while being accelerated outward. Thus, in this more
detailed picture of the UFO-host ISM interaction one may expect some cold gas
streaming outward in the bulge at velocities consistent with the
energy-conserving flow, consistent with the observations. Further developments
in the field should utilise numerical simulations to resolve the multi-phase
ISM shocked by the UFOs in more realistic detail.

\section{Conclusions}

Here we attempted to extend the model of \cite{King03} for AGN feedback on the
case when the ambient ISM in the host has a multi-phase structure. In our
two-phase approximation, the ISM consists of a cold dense phase, such as cold
clouds, and a hot tenuous phase that fills most of the volume. We proposed
that the UFO overtakes the cold clouds, as supernova shock waves do. The
clouds are affected by a strong but nearly isotropic compression due to the
reverse shock in the UFO, and a smaller outward push. The interaction of the
UFO with the hot ISM, on the other hand, is much closer to the spherically
symmetric shock picture employed in previous work
\citep[e.g.,][]{King10b,ZubovasEtal11a}. This phase is blown away from the
bulge by the UFO easily but is constantly replenished by cloud ablation and
destruction.

We showed that a combination of constraints from the outward push on the cold
ISM and cloud destruction by star formation sets an $M-\sigma$ relation
similar in form to that derived by \cite{King03}. Main model predictions
distinguishing it from previous work are (1) the importance of triggered star
formation in the cold gas in limiting SMBH growth and (2) the ability of
``underweight'' black holes to affect their hosts in the energy conserving
mode due to percolation of hot gas to large distances via low density
voids. 

We believe that SMBH-host galaxy connections cannot be properly understood
without taking into account the multi-phase structure of the ISM of the host.

\section*{Acknowledgments}

Theoretical astrophysics research in Leicester is supported by an STFC
grant. The author is indebted to Andrew King for a number of discussions and
comments on the paper. Ken Pounds, Kastytis Zubovas and Martin Bourne are also
thanked for many useful discussions.  This research used the ALICE High
Performance Computing Facility at the University of Leicester, and the DiRAC
Facility jointly funded by STFC and the Large Facilities Capital Fund of BIS.

\bibliographystyle{mnras} 

\label{lastpage}

\end{document}